# Fine-tuning universal machine learning potentials for transition state search in surface catalysis

Raffaele Cheula[1,2]*, Mie Andersen[2]*, and John R. Kitchin[1]*

[1] Department of Chemical Engineering, Carnegie Mellon University, Pittsburgh, Pennsylvania 15213, United States

[2] Center for Interstellar Catalysis, Department of Physics and Astronomy, Aarhus University, 8000 Aarhus C, Denmark

*Corresponding authors: raffaele.cheula@phys.au.dk, mie@phys.au.dk, jkitchin@andrew.cmu.edu

**Abstract**

Determining transition states (TSs) of surface reactions is central to understanding and designing heterogeneous catalysts but remains computationally prohibitive with density functional theory (DFT). While machine learning potentials (MLPs) offer significant speedups, task-specific models have limited transferability across catalytic systems, and universal MLPs (uMLPs) lack the accuracy needed for reactive configurations. Here, we present a workflow based on active learning to iteratively fine-tune uMLPs for DFT-quality TS search. Using 250 TSs from the $CO_2$ hydrogenation reaction network on metal and single-atom alloy surfaces, we first benchmark TS search algorithms, identifying the Sella algorithm as most robust, and propose a modification (Bond-Aware Sella) that substantially improves its success rate. We then explore sequential and batch active-learning strategies for fine-tuning and show that DFT-quality TS structures can be found using only 8 DFT single-point calculations on average per structure. This demonstrates the viability of fine-tuned uMLPs for high-throughput catalyst screening.

**Introduction**

Metals and alloys are important catalyst materials for various industrial chemical processes within energy conversion, emissions control, and chemical synthesis. Understanding their activity or selectivity to desired products, or designing improved catalyst materials, requires understanding of the reaction mechanism at the atomic level[1–6]. Central to this task is the ability to efficiently determine the structure and energy of the TS of elementary reaction steps, i.e., the highest-energy structure along the minimum energy path connecting reactants to products on the potential energy surface (PES). Mathematically, TSs are located by finding first-order saddle points on the PES, i.e., points having zero gradient and exactly one negative eigenvalue in the Hessian. However, identifying these elusive saddle points is a very complex computational challenge, especially in surface catalysis, where the PES is, in general, high-dimensional. Several algorithms have been developed for TS search over the years, including the Nudged Elastic Band (NEB) method[7], the Dimer method[8,9], the Automated Relaxed Potential Energy Surface Scans (ARPESS) method[10], and the Sella algorithm[11]. All these TS search algorithms are computationally demanding when DFT methods are employed for calculating the energies and forces of the structures.

A significant speedup, often several orders of magnitude, can be obtained when replacing DFT with an MLP. MLPs trained to appropriately chosen DFT structures relevant to the task at hand may reach near-DFT accuracy, even for complex tasks like finding TSs of surface reactions[12–14]. Unfortunately, such task-specific MLPs have very limited transferability to other systems of interest, e.g., a new metal, alloy, or elementary reaction step. This makes them of little use in catalyst design studies, which typically aim to span across a significant portion of metals in the periodic table. On the other hand, recently developed universal MLPs (uMLPs), e.g., CHGNet[15], MACE-MP-0[16], or models from the Open Catalyst Project (OCP)[17–19], are pre-trained to millions of structures covering elements from the entire periodic table. As such, they are very promising for catalyst design, although they generally lack the accuracy of task-specific models. uMLPs have shown impressive zero-shot performance for equilibrium materials properties like phonons[20] and for accelerating DFT-based TS search in surface catalysis through pre-optimization[21]. Strategies to fine-tune uMLPs for problematic cases, such as systematic softening, which occurs due to biased sampling of near-equilibrium structures in data sets generated for uMLP training, have also been proposed[22]. For surface catalysis, uMLPs from the OCP trained on adsorbate-covered pure metal and intermetallic surfaces have been fine-tuned to predict adsorption energies at high-entropy alloy surfaces[23].

In this work, we introduce a workflow for fine-tuning uMLPs for TS search in surface catalysis. We consider 250 TSs for elementary steps in the reverse water-gas shift reaction network ($CO_2$ hydrogenation to CO and $H_2O$) on metals and single-atom alloy surfaces[24]. First, we exploit the low cost of uMLPs (without fine-tuning) to benchmark literature single-ended TS search algorithms, finding the Sella algorithm to be the most robust. We further propose a modification to the Sella algorithm (Bonds-Aware Sella), which substantially increases its success rate for finding the intended TS. Using this modified algorithm, we explore two different iterative active-learning strategies (sequential and batch) for fine-tuning uMLPs on structures encountered during TS search. We show that the best strategy (the sequential) allows for obtaining DFT-quality TS structures with only 8 iterations on average. This corresponds to a reduction in cost from ~100 DFT single-point calculations per structure required on average for DFT-only TS optimization to 8 DFT single-points on average in the fine-tuning workflow. We expect that the developed workflow may help accelerate various aspects of catalysis research, such as elucidating the microscopic reaction mechanism behind experimentally observed active catalysts or high-throughput computational catalyst screening to find new catalysts.

## Results

*Bonds-Aware Sella method*

Locating TSs of elementary reactions remains a challenging task: double-ended methods like NEB typically require hundreds of optimization steps over multiple images along the reaction coordinate connecting the initial and final state, resulting in thousands of forces evaluations[21]. On the other hand, single-ended TS search methods that consider only one or two images, such as the Dimer method and Sella, usually reduce substantially the number of required forces evaluations, but sometimes fail to identify the intended TS structure[25]. Here we exploit the fact that, for surface reactions in catalysis, the bonds expected to form and break during the elementary step are usually known *a priori* by construction of the reaction pathway. This valuable information can be exploited to guide the TS search. Building on this observation, we incorporate bond-formation and bond-breaking information directly into the Sella optimization scheme to enhance its robustness, resulting in the Bonds-Aware (BA) Sella method, illustrated in Fig. 1.

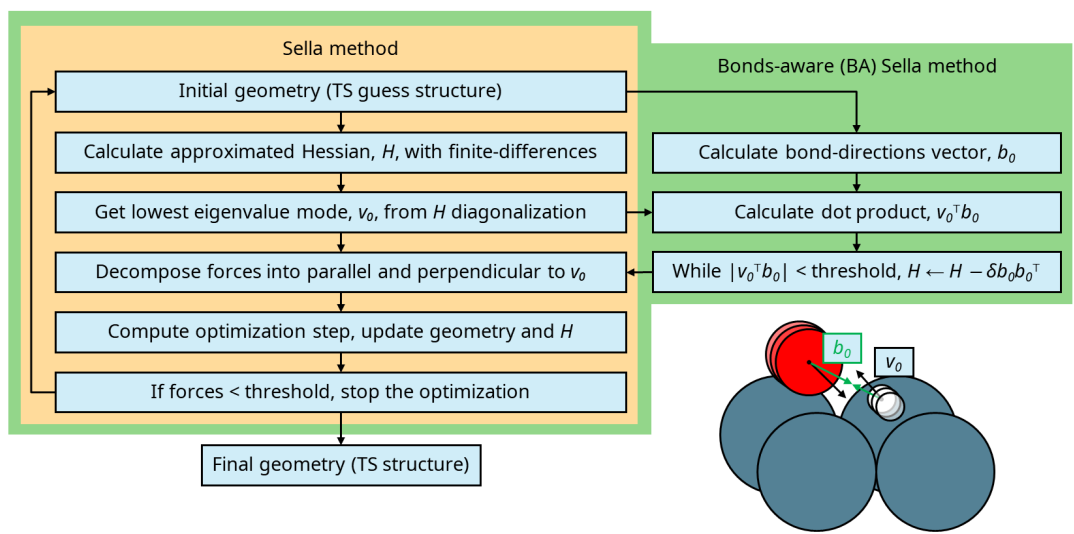

**Figure 1**: **Workflow of the Sella and Bonds-Aware Sella (BA-Sella) transition-state optimization methods**. The standard Sella algorithm (left) approximates the Hessian, identifies the lowest-eigenvalue mode, and updates geometry and approximated Hessian. BA-Sella (right) augments this by introducing a bond-direction vector that checks alignment with the reaction coordinate and adjusts the approximate Hessian when needed, guiding the search toward the intended TS.

In the standard Sella[11] approach, the optimization begins from an initial TS guess, followed by a finite-difference curvature evaluation to construct the approximate Hessian, $H$. The lowest-eigenvalue mode, $v_0$, is then obtained through iterative diagonalization of $H$, and the gradient (forces acting on atoms) is decomposed into components parallel and perpendicular to $v_0$. An optimization step is constructed using restricted-step partitioned rational function optimization (RS-PRFO), in which the system is driven uphill along $v_0$ and downhill in the orthogonal space. The step length is controlled within a trust-radius framework before updating geometry and $H$. This procedure is repeated until the maximum force falls below a predefined threshold.

The BA extension builds on this procedure by introducing a chemically informed bond-direction vector, $b_0$, which describes expected bond-breaking or bond-forming motions. At each iteration, the alignment between $v_0$ and $b_0$ is quantified by the absolute dot product, $|v_0^T b_0|$, with both vectors normalized to unit length. If this alignment falls below a chosen threshold, $H$ is selectively modified by repeatedly subtracting $\delta b_0 b_0^T$ (rank-one updates), which decreases the curvature along the chemically expected $b_0$ direction until the lowest-curvature eigenmode aligns with that direction ($|v_0^T b_0|$ reaches the threshold value). This integration of information helps guide the optimizer toward the intended reaction pathway, improving robustness and reducing failures associated with misleading curvature modes near the TS region.

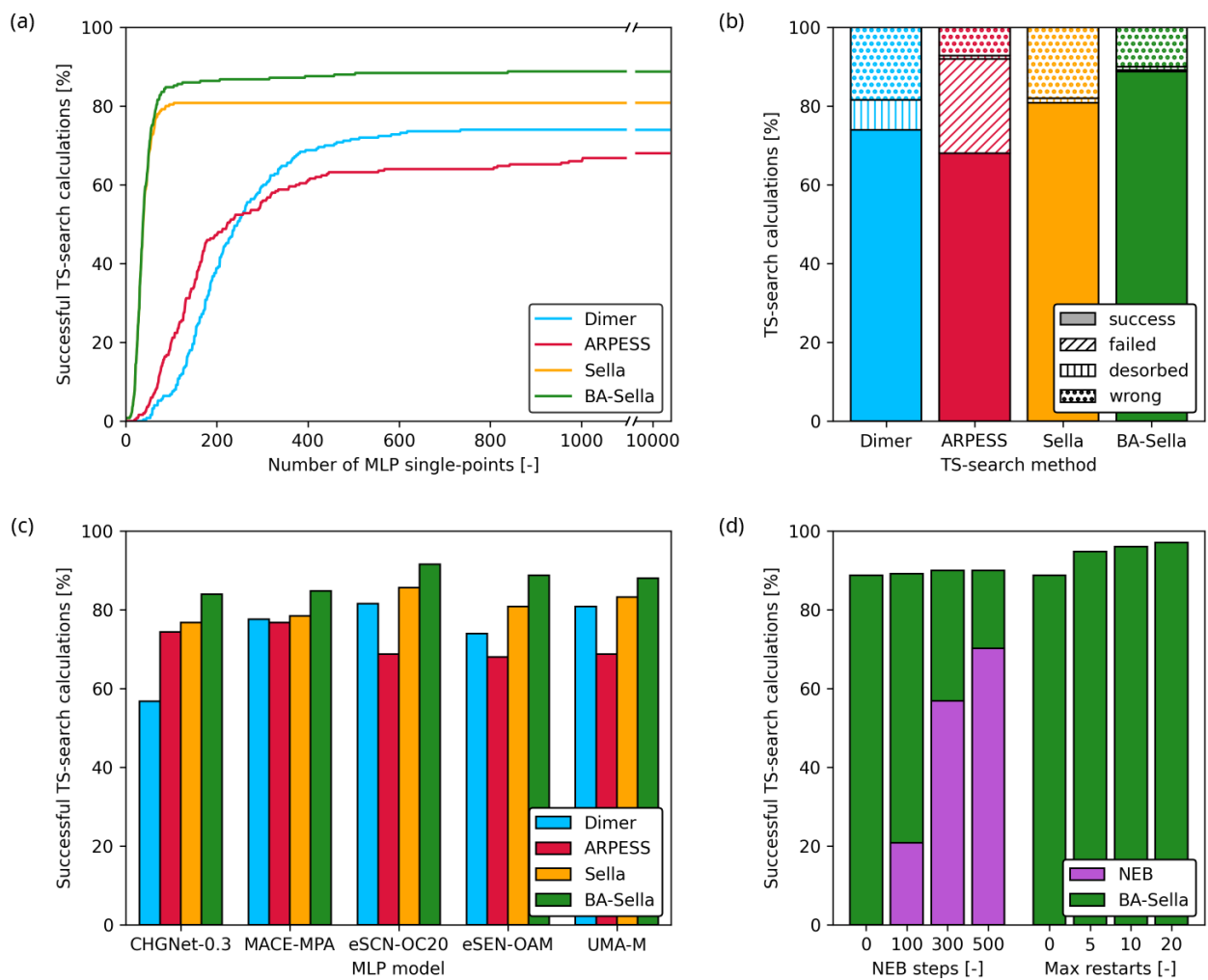

**Figure 2**: **Performance of TS-search algorithms using MLP single-point evaluations.** (a) Cumulative success rate of TS searches as a function of the number of MLP single-point evaluations, obtained with the eSEN-OAM potential. (b) Distribution of TS-search outcomes using eSEN-OAM, distinguishing successful optimizations ("success") from unsuccessful cases ("desorbed", "failed", and "wrong"). (c) TS-search success rates for different MLPs and TS-search methods. (d) Success rates obtained with BA-Sella and eSEN-OAM when (left) preceded by a given number of NEB steps or (right) combined with multiple stochastic restarts.

*Performance of TS-search methods*

To address the efficiency and robustness of the Bonds-Aware Sella (BA-Sella) method, we compare its performance to three other TS search algorithms: Dimer[8], ARPESS[10], and the original Sella algorithm[11]. We test the four TS search algorithms for 250 TSs for elementary steps in the reverse water-gas shift reaction network, for which we previously performed NEB TS optimizations with DFT[24]. Here, the eSEN-OAM[26] pre-trained uMLP (see Methods) is used for all optimizations. The initial TS guess is constructed in the same way for all tested algorithms (see Methods). Figure 2a shows the cumulative percentage of successful TS searches as a function of the number of MLP single-point evaluations. We adopt this metric, rather than the number of optimization steps, because different algorithms require a different number of single-point force evaluations per step. A calculation is considered successful if it locates a TS structure corresponding to the intended elementary step (no comparison with DFT references is made at this stage). Sella and BA-Sella converge markedly faster than the other methods, and BA-Sella reaches the highest overall success rate (88%), while standard Sella plateaus at approximately 80%. The Dimer method requires substantially more evaluations to reach slightly lower success levels (74%), and ARPESS exhibits the slowest and lowest success overall (68%). Figure 2b summarizes the outcomes of the TS searches. Successful calculations, which identified the intended TS structure, are indicated as "success", while unsuccessful calculations are distinguished between "desorbed" (in which the atoms of the TS structure are desorbed from the surface), "failed" (in which the optimization algorithm did not converge or encountered termination errors), and "wrong" (in which the TS-search algorithm did not locate the intended TS structure, but a minimum in the PES or another TS). We verify if the output TS structure corresponds to the intended elementary step by calculating the TS mode with finite differences and relaxing the structure after displacement in the two directions of the TS mode; if the chemical bonds between the atoms in the reacting molecules (not surface atoms) do not correspond to the ones of the initial or final states (one for each direction), the calculation is considered "wrong". BA-Sella exhibits the largest fraction of correctly converged TSs with minimal desorption or failure events, whereas ARPESS shows a significant proportion of incorrect or failed searches. These results collectively demonstrate that incorporating chemically informed bond-direction guidance into Sella improves both the robustness and efficiency of TS optimization compared to existing methods.

To evaluate the impact of the underlying uMLP, we repeat the TS-search calculations using four other state-of-the-art uMLPs and compare their performance (Figure 2c). These models are CHGNet-0.3[15] (trained on MPTrj[15,27]), MACE-MPA[16] (trained on MPTrj[15,27] and sAlex[28,29]), eSCN-OC20[30] (trained on OC20[17]), and UMA-M[31] (trained on OMol25[32], OMat24[29], OC20[17], ODAC25[33], and OMC25[34]), together with the previously discussed eSEN-OAM[26] (trained on MPTrj[15,27], sAlex[28,29], and Omat24[29]). More details on these uMLPs are given in the Methods section. Across all uMLPs, BA-Sella consistently achieves the highest success rate, followed by the original Sella algorithm. The three FAIRChem models (eSCN-OC20, eSEN-OAM, and UMA-M) yield comparable results, with BA-Sella success rates between 87% and 90%. CHGNet-0.3 and MACE-MPA show slightly lower BA-Sella success rates (~84%). These results indicate that the improvements introduced by BA-Sella are largely independent of the specific MLP employed, although the overall success rate remains influenced by the model architecture and training data. The cumulative success rate of TS searches and the distribution of TS-search outcomes for these uMLPs are reported in the Supplementary Information, Figure S1-S4.

We then explore strategies to further enhance TS-search success rates (Figure 2d). First, we investigate whether improving the initial TS guess through preliminary NEB calculations increases the overall success rate. The NEB calculations themselves converge slowly: even after 500 NEB steps, only ~70% of the calculations converged to a TS. This highlights the difficulty and computational cost of NEB for surface reactions. Moreover, when BA-Sella is applied starting from the NEB-refined TS guess, the overall success rate (counting TSs found either by NEB or BA-Sella) does not increase significantly. This suggests that the limiting factor is not the quality of the initial geometry. We also examine a stochastic-restart strategy, in which unsuccessful calculations are restarted after applying random perturbations to both the TS guess structure and the optimizer settings. Specifically, we apply a collective random rigid translation and individual random displacements to the atoms of the reacting molecules, together with quasi-random perturbations of selected optimization parameters (see Methods). This approach substantially improves robustness: with increasing numbers of maximum restarts, the BA-Sella success rate steadily increases, reaching ~97% when 20 restarts are allowed. Overall, these results demonstrate that chemically informed curvature control combined with controlled stochastic restarts provides a highly robust framework for TS optimization of complex surface reactions.

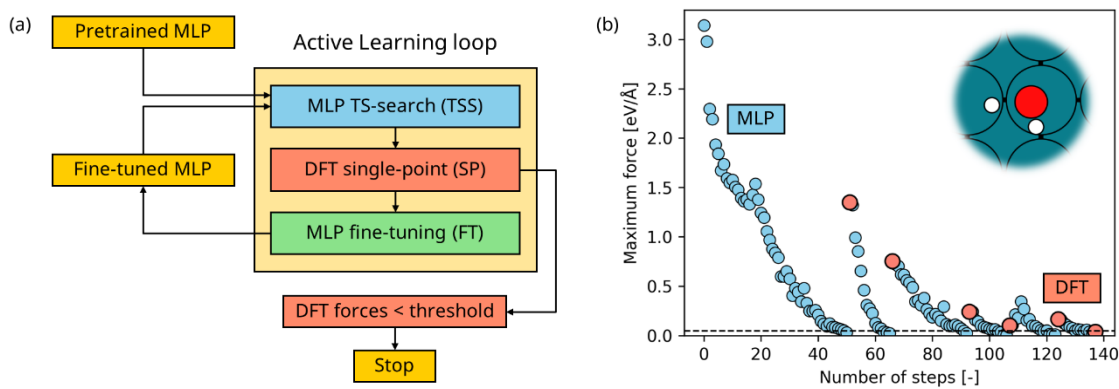

**Figure 3**: **Iterative fine-tuning of the uMLP.** (a) Workflow of the MLP refinement cycle, where MLP-based optimizations (TS search) are followed by DFT single-point evaluations and subsequent MLP fine-tuning until the DFT forces fall below a chosen threshold. (b) Evolution of the forces during the procedure, showing the progressive reduction in both MLP-predicted forces (blue points) and DFT forces (red points).

*Active learning strategies*

With the identified best-performing BA-Sella TS search method, we proceed to test different active learning-based fine-tuning approaches and address their efficiency in identifying TS structures with DFT accuracy. A general scheme of the active learning workflow is reported in Figure 3a. The workflow begins with a pretrained MLP that is used to optimize the TS, terminating when the forces reach a user-defined threshold value. A DFT single-point calculation is then performed on the optimized TS structure, and the resulting forces are used to fine-tune the MLP. This loop (MLP optimization, DFT evaluation, and MLP retraining) is repeated until the DFT forces fall below a defined threshold, yielding the geometry and energy of the TS with DFT accuracy. Figure 3b shows how both the MLP-predicted forces and the DFT forces decrease during the procedure, illustrating that each refinement step brings the MLP predictions closer to DFT accuracy.

We test two different active learning strategies. In the sequential active learning approach (Figure 4c), each calculation is treated independently: an MLP is iteratively fine-tuned using DFT single-point evaluations generated along the optimization trajectory and then used to further optimize the same structure, yielding a highly specialized potential for each individual calculation. In contrast, in the batch active learning approach (Figure 4d), the active learning loop is carried out simultaneously across different TS searches, and the DFT data collected from all trajectories are combined to fine-tune a single MLP that is reused for all calculations, resulting in a more general and transferable potential. These active learning approaches are compared to two simpler strategies. In the first approach (Figure 4a), TS structures are optimized entirely using DFT forces, starting from the initial TS guess. In the second approach (Figure 4b), TS structures are first pre-optimized with the MLP and subsequently refined with a full DFT-based optimization.

Figure 4e compares the distributions of the number of DFT single-point calculations required by the four approaches, using the eSEN-OAM uMLP. The full DFT optimization shows the highest computational cost, with a mean of ~102 DFT single-point evaluations per structure and a broad distribution characterized by a long tail extending to several hundred calculations. The DFT optimization after pre-optimization with an MLP reduces the computational effort to a mean of ~70 DFT single-point evaluations, although a substantial spread remains. The sequential active learning strategy provides the most efficient performance, requiring on average only ~8 DFT single-point calculations per structure, with most cases tightly clustered at low values. The batch active learning approach also leads to a significant reduction in DFT single-points compared to the DFT-based workflows, with a mean of ~38 DFT single-point evaluations but exhibits larger variability than the sequential approach. Overall, both active learning strategies dramatically decrease the number of expensive DFT evaluations, with sequential active learning achieving the greatest and most consistent reduction.

The superior efficiency observed for the sequential active learning approach can be attributed to the strong specialization of the MLP to each individual target structure. In this strategy, the MLP is repeatedly fine-tuned using DFT data generated along a single optimization trajectory, which likely leads to overfitting toward the specific local region of the PES explored by that structure. As a result, the fine-tuned model is not transferable and cannot be reliably applied to other systems. Nevertheless, this strong specialization is advantageous in terms of computational efficiency, as it yields the highest speedup and the lowest number of required DFT single-point calculations per structure. In contrast, the batch active learning approach aggregates DFT data from multiple trajectories to fine-tune a single MLP, producing a more general and transferable model that can be reused for further applications, at the cost of a higher average number of DFT calculations per structure.

Figure 4f summarizes the distribution of TS-search outcomes for the four workflows, classified as "matched" if the same TS as in the original NEB dataset is identified (with a difference in energy lower than 0.1 eV), or "lower" if a TS structure with lower energy is found. Unsuccessful outcomes are denoted "higher" (if a TS with higher energy is found), "failed" (if the calculation has not converged to a TS structure), and "missing" (if the initial MLP TS search did not identify any TS structure). The maximum number of DFT single-points is set to 500 for the optimizations with DFT and 250 for the active learning approaches, as we observed that the computational time of one DFT single-point is similar to one uMLP fine-tuning (see Figure S5). The two optimizations with DFT exhibit high fractions of correctly matched TS structures but also show non-negligible proportions of higher-energy or failed cases, reflecting the intrinsic difficulty of TS optimization. The sequential active learning approach achieves a similar, even slightly higher, fraction of matched structures compared to the DFT approach, demonstrating that the reduction in computational cost (Figure 4e) does not sacrifice robustness in converging to the intended TS. The batch active learning strategy also maintains a high success rate, although it displays a slightly increased fraction of higher-energy or mismatched outcomes compared to the sequential approach. Overall, both active learning schemes preserve DFT-level reliability while substantially decreasing the number of expensive DFT evaluations, with sequential active learning providing the most consistent performance across the tested reactions.

These results become even more compelling when benchmarked against the original NEB dataset used to generate the reference TS structures. In that dataset, each NEB calculation required, on average, ~250 optimization steps. Considering that each NEB step involves force evaluations across multiple images, this corresponds to roughly ~2000 DFT single-point calculations per TS search. In this context, even the DFT single-structure optimizations presented here are already substantially more efficient, while the active learning strategies, particularly the sequential approach, reduce the computational cost by two to three orders of magnitude. This dramatic decrease in computational cost highlights the potential of combining single-structure TS-search algorithms with active-learning workflows for exploration of large-scale reaction networks and catalyst design.

We also evaluated a simpler workflow in which the TS searches are performed using the uMLP, followed by DFT single-point calculations on the MLP-optimized TS structures. As shown in Figure S6, this approach identifies a TS matching the reference NEB structure (with a difference in energy lower than 0.1 eV) in approximately 70% of the cases, consistent with previously reported results[21]. Although this strategy results in reduced accuracy in TS energy predictions compared to DFT or active learning approaches, it provides an efficient low-cost screening step prior to more rigorous refinement.

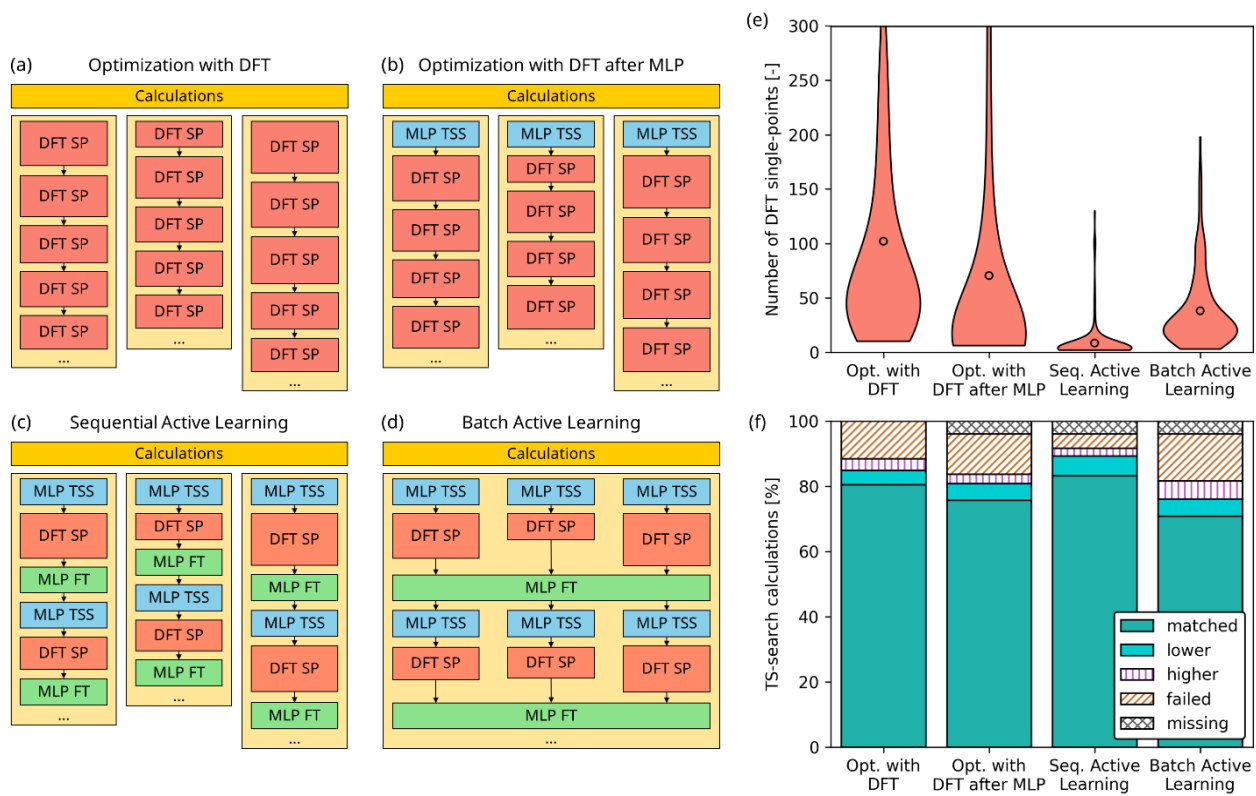

**Figure 4**: **Schematic representation of the four computational workflows and their corresponding computational cost in terms of DFT single-point evaluations.** The approaches include: (a) optimization with DFT, (b) optimization with DFT after MLP pre-optimization, (c) sequential active learning, where each structure is treated independently, and (d) batch active learning, where DFT data from multiple structures are aggregated to fine-tune a single, more general MLP reused across all calculations. The violin plot (e) shows the distribution of the number of DFT single-point calculations required for each method. Mean values are indicated by black markers. (f) Outcomes of the TS searches for each workflow, classified as "matched" (correct TS identified), "lower" (convergence to a lower-energy TS), "higher" (convergence to a higher-energy TS), "failed" (no convergence), and "missing" (no TS structure produced by the MLP).

**Discussion**

This work demonstrates that combining chemically informed TS search strategies with sequential fine-tuning of uMLPs provides a practical and scalable framework for automated TS discovery. By incorporating bond-aware curvature control into the Sella optimizer, the BA-Sella method introduced here achieves both high efficiency and strong robustness compared to state-of-the-art TS-search approaches for surface reactions. In practice, it requires a small number of single-point evaluations to find a TS structure while maintaining a high success rate. Furthermore, its single-image formulation facilitates seamless integration into active-learning workflows involving MLP fine-tuning. This compatibility enables the location of TSs with DFT-level accuracy while substantially reducing the number of costly DFT evaluations.

The ability to identify TS structures of complex surface reactions with an average of less than 10 DFT single-point calculations represents a major step forward in computational catalysis. Conventional NEB calculations typically require on the order of thousands of DFT single-point evaluations per elementary step. In comparison, the present approach reduces the computational effort by two to three orders of magnitude. Even relative to direct single-ended DFT-based TS optimizations, the active learning workflows substantially lower the cost while maintaining high reliability. This reduction significantly expands what is practically feasible in first-principles mechanistic studies.

Among the tested strategies, sequential active learning clearly delivers the best overall performance. It combines the lowest computational cost with the most reliable convergence to the intended saddle points. By repeatedly refining the MLP along a single optimization trajectory, the model quickly adapts to the specific region of the PES relevant to the elementary step under study. Although this strong specialization limits transferability to other systems, it maximizes efficiency for individual TS searches. In contrast, the batch strategy, while still more efficient than purely DFT-based workflows, requires more DFT single-point calculations and shows larger variability because the model must remain more general across different structures.

We need to stress that the present validation has been carried out for surface elementary reactions, where the reaction coordinate can be described in terms of a limited number of bond-forming and bond-breaking events. The approach is therefore expected to extend naturally to molecular reactions, which have similar localized chemical characters. However, solid-state transformations (e.g., surface reconstructions, bulk phase transitions, or collective lattice distortions) are more complex and cannot always be described as a small set of bond rearrangements. Applying chemically informed search directions to such collective processes will therefore likely require further methodological developments.

The strong reduction in computational cost enabled by the active-learning strategies described here makes TS searches possible at a scale that was previously unrealistic. In heterogeneous catalysis, understanding reaction mechanisms often requires exploring large reaction networks across multiple surfaces, active sites, and coverages to establish reliable microkinetic models. Being able to compute accurate activation barriers with minimal DFT effort opens the door to systematic, high-throughput evaluation of elementary steps and more predictive kinetic models based on first-principles calculations.

## Methods

*Atomistic optimizations*

Atomistic structure optimizations were performed using the Atomic Simulation Environment (ASE) library[35]. Relaxations of initial and final states of elementary steps, as well as NEB calculations, were carried out using the Broyden-Fletcher-Goldfarb-Shanno (BFGS) algorithm. NEB calculations employed 10 images. The climbing-image (CI) scheme was activated once the maximum force dropped below 0.10 eV Å$^{-1}$. All single-ended TS searches were performed through their ASE implementations (ARPESS corresponds to ClimbFixInternals in ASE). Unless otherwise specified, the force convergence threshold was set to 0.05 eV Å$^{-1}$. Vibrational analyses used to calculate TS modes were performed by finite differences. The parameters of all optimizers used are reported in the Supplementary Information.

*BA-Sella method*

In the BA-Sella method, the bond direction vector, $b_0$, was constructed by summing the normalized vectors connecting atoms involved in bonds expected to form or break in the TS mode. These bonds were identified by comparing the initial and final states of the elementary step and selecting the bonds connecting the atoms of the reacting molecules that change connectivity between the two states (bonds involving surface atoms were not considered). For forming bonds, the sign of the two vectors was chosen so that they point along decreasing interatomic distance (atoms moving closer). For breaking bonds, the sign was reversed, corresponding to increasing interatomic distance (atoms moving apart). This sign convention ensures that, when multiple bonds are involved, their contributions add consistently and $b_0$ represents the expected collective TS mode. The threshold value of the dot product, $|v_0^T b_0|$, was set to 0.5.

*Initial TS guess generation*

The initial TS guess structures for single-ended TS-search algorithms were obtained by interpolating between the initial and final states of the elementary steps with the image-dependent pair potential (IDPP) method[36], using 10 images, a force convergence threshold of 0.01 eV Å$^{-1}$, and a maximum of 1000 steps. The image with the highest energy (evaluated with the method of interest: DFT or MLP) among the eight intermediate images (excluding the initial and final states) was selected as the TS guess. This procedure provides a physically reasonable starting TS structure while requiring only eight energy evaluations.

*Stochastic-restart strategy*

In the stochastic-restart strategy, atomic positions were perturbed using random numbers sampled from normal (Gaussian) distributions. Specifically, we applied a collective rigid translation and individual displacements to the atoms of the reacting molecules. The distributions for sampling translation and displacements were centered at zero, with standard deviations of 0.25 Å and 0.15 Å, respectively. In addition, BA-Sella optimizer parameters were perturbed using a Sobol quasi-random sequence (SciPy[37] implementation). The perturbation ranges are provided in the Supplementary Information.

*Universal MLPs*

The following uMLPs were used, through their ASE interface: CHGNet-0.3 (CHGNet[15] version 0.3.0), trained on the MPTrj[15,27] dataset, with ~400k parameters; MACE-MPA (MACE-MPA-0[16]), trained on MPTrj[15,27] and sAlex[28,29] (subsampled Alexandria dataset), with ~9M parameters; eSCN-OC20 (eSCN-L6-M3-Lay20-S2EF-OC20-All+MD[30]), trained on the OC20[17] dataset, with ~200M parameters; eSEN-OAM (eSEN-30M-OAM[26]), trained on MPTrj[15,27], sAlex[28,29], and OMat24[29] datasets, with ~30M parameters; UMA-M (UMA-medium[31], OC20 task), trained OMol25[34], OMat24[29], OC20[17], ODAC25[33], and OMC25[34], using a mixture of linear experts, with 1.4B parameters (but only ~50M active parameters per atomic structure).

*Fine-tuning and active learning*

Fine-tuning of the eSEN-OAM model was performed for 100 epochs using a cosine learning-rate scheduler and a low learning rate ($10^{-4}$). Only forces were used during fine-tuning; energies were excluded[38]. At each fine-tuning iteration, a fresh pre-trained model was initialized and fine-tuned, discarding previously fine-tuned weights to prevent the progressive degradation of the underlying learned representation. Fine-tuning was performed using only the final structure obtained from the MLP-driven TS search. Within the active-learning loops, TS optimizations were conducted with a tighter force convergence threshold (0.01 eV Å$^{-1}$) and a minimum of 10 optimization steps.

*DFT calculations*

DFT calculations were performed with the Quantum Espresso[39–41] suite of codes, using the same settings as in the work by Cheula et al.[24], i.e., the BEEF-vdW exchange-correlation functional[42], pseudopotentials from the SSSP library[43], and a plane wave basis set. The plane wave and electronic density cut-off energies were set to 40 Ry (~544 eV) and 320 Ry (~4354 eV), respectively. The convergence threshold selected for the electronic energy of self-consistent field (SCF) steps was $10^{-6}$ Ry (~1.36 $10^{-5}$ eV). Efficient communication between ASE and Quantum ESPRESSO was ensured through an ASE socket calculator using the i-PI[44] protocol.

**Data availability**

The data generated and analyzed during this study are available at: https://github.com/raffaelecheula/arkimede in the form of ASE databases of atomic structures.

**Code availability**

The code for running the TS-search calculations, including the BA-Sella algorithm and the active-learning schemes, is accessible at: https://github.com/raffaelecheula/arkimede. The code to fine-tune the machine learning potentials used in this work is available at: https://github.com/raffaelecheula/mlps_finetuning.

**Acknowledgments**

RC acknowledges support from European Union's Horizon Europe research and innovation programme under the Marie Skłodowska-Curie grant no. 101108769. MA acknowledges funding from Novo Nordisk Fonden (grant no. NNF22OC0078939), VILLUM FONDEN (grant no. 37381), and the Danish National Research Foundation through the Center of Excellence "InterCat" (grant no. DNRF150). Computational support was provided by the Centre for Scientific Computing Aarhus (CSCAA) at Aarhus University.

**Authors contributions**

RC conceptualized the study, conducted the research, and wrote the manuscript. MA and JK contributed to the conceptualization of the study, supervised the research, and edited the manuscript. RC and MA acquired funding. All authors discussed the results and approved the final manuscript.

**Competing Interests**

All authors declare no financial or non-financial competing interests.

# Supplementary Information: Fine-tuning universal machine learning potentials for transition state search in surface catalysis

Raffaele Cheula[1,2]*, Mie Andersen[2]*, and John Kitchin[1]*

[1] *Department of Chemical Engineering, Carnegie Mellon University, Pittsburgh, Pennsylvania 15213, United States*

[2] *Center for Interstellar Catalysis, Department of Physics and Astronomy, Aarhus University, 8000 Aarhus C, Denmark*

*Corresponding authors: raffaele.cheula@phys.au.dk, mie@phys.au.dk, jkitchin@andrew.cmu.edu

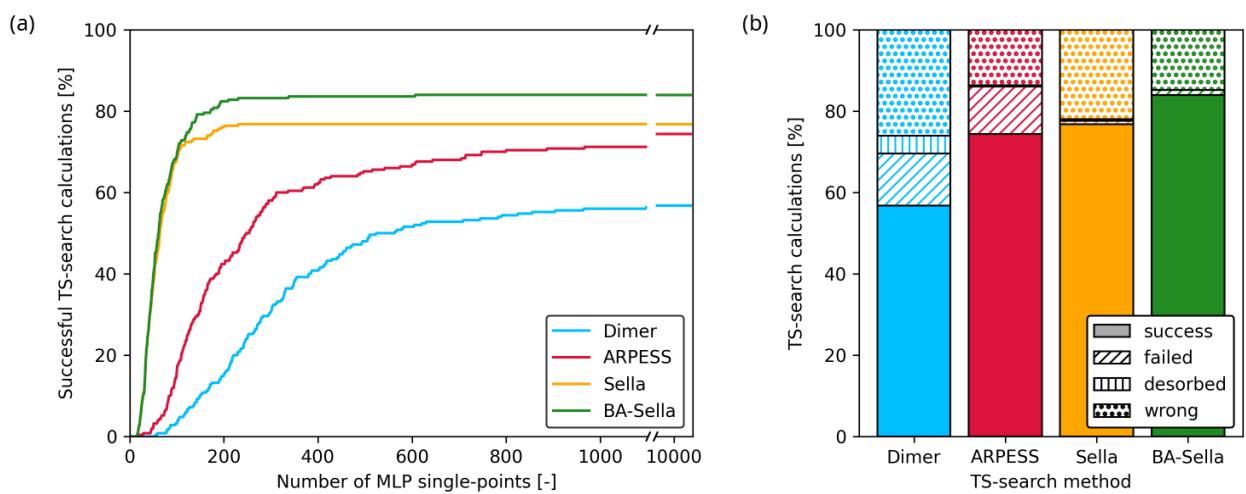

**Figure S1**: (a) cumulative success rate of TS searches and (b) distribution of TS-search outcomes obtained with CHGNet-0.3.

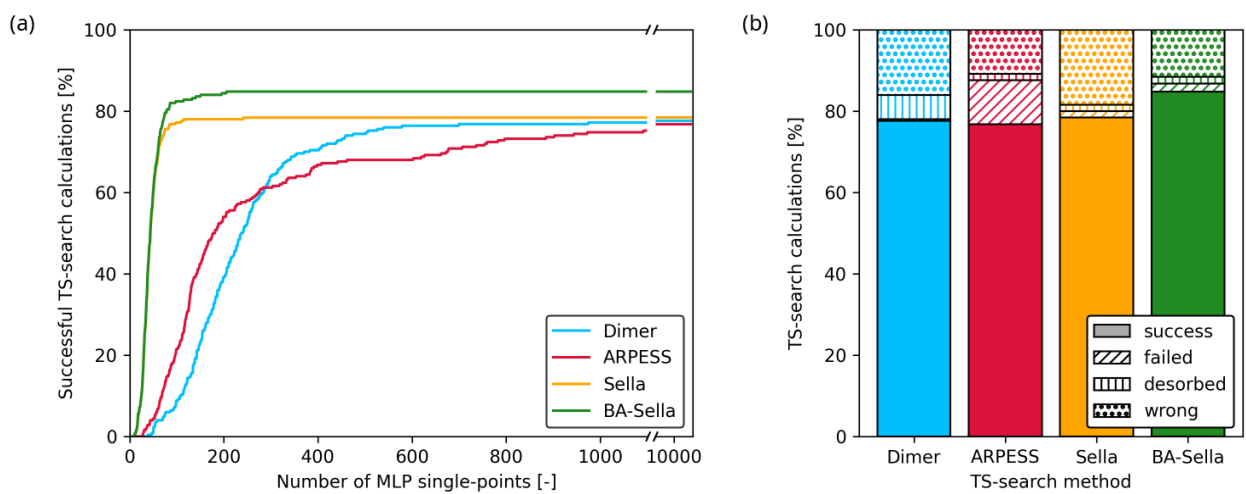

**Figure S2**: (a) cumulative success rate of TS searches and (b) distribution of TS-search outcomes obtained with MACE-MPA.

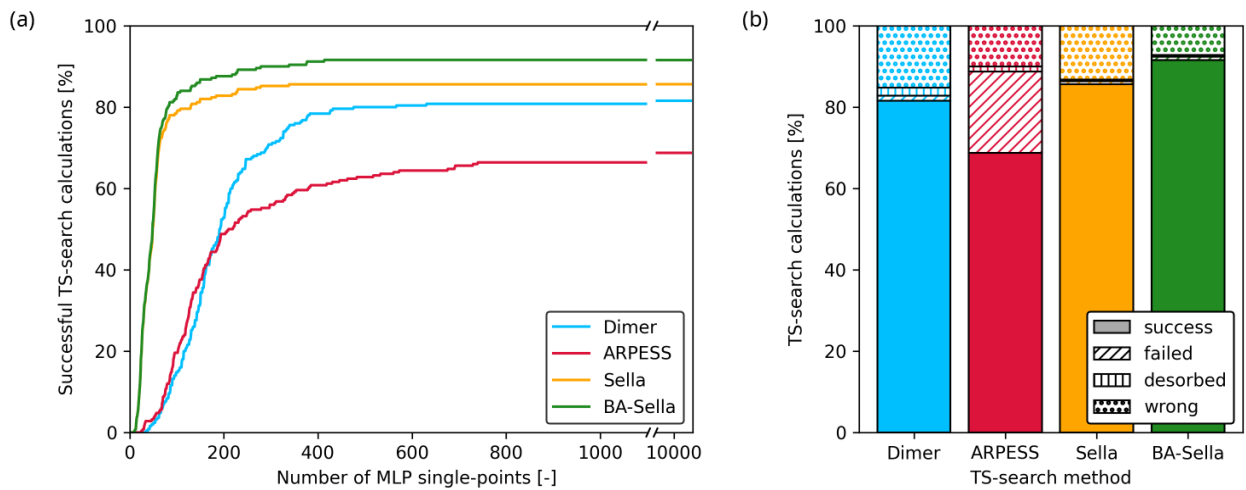

**Figure S3**: (a) cumulative success rate of TS searches and (b) distribution of TS-search outcomes obtained with eSCN-OC20.

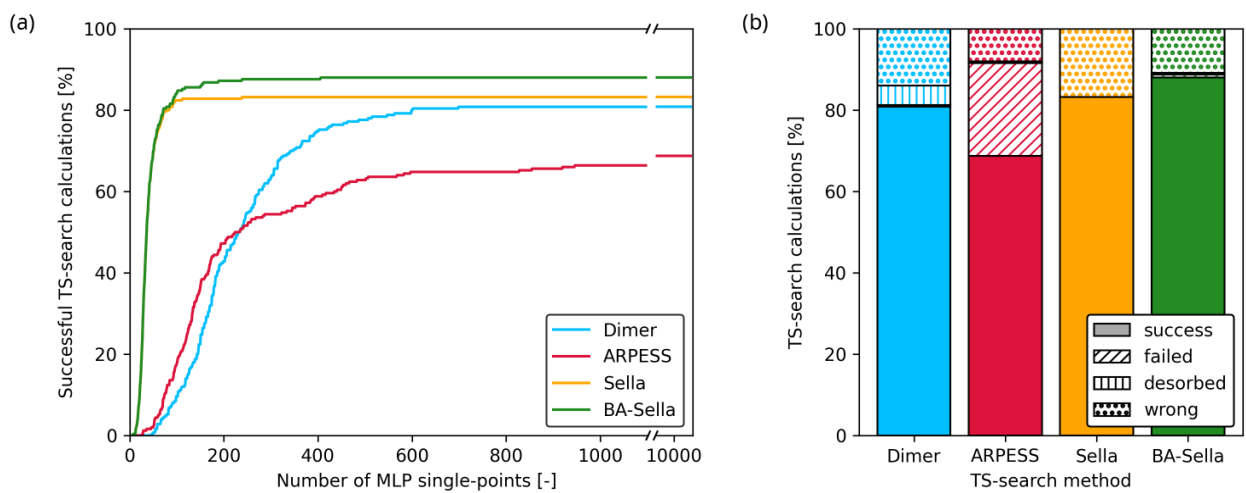

**Figure S4**: (a) cumulative success rate of TS searches and (b) distribution of TS-search outcomes obtained with UMA-M.

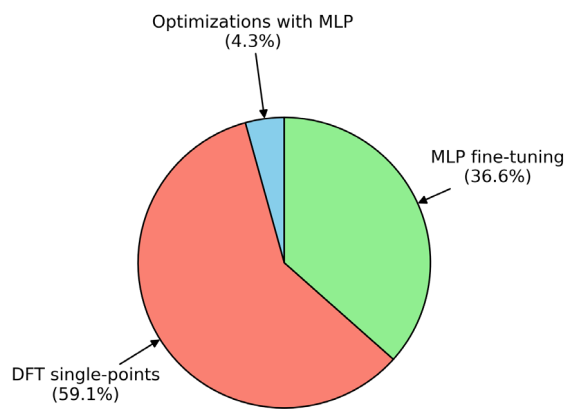

**Figure S5**: Average computational times of optimizations (TS searches) with MLP, DFT single-points, and MLP fine-tuning, during sequential active learning with eSEN-OAM.

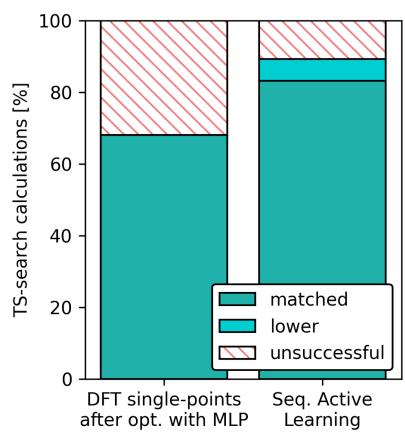

**Figure S6**: Outcomes of the TS searches for two workflows: TS searches with uMLP followed by DFT single-point calculations, and sequential active learning. Outcomes are classified as "matched" (correct TS identified), "lower" (convergence to a lower-energy TS), or "unsuccessful" (unsuccessful identification of the correct TS structure or a lower-energy TS corresponding to the starting reaction step).

# Code Guide: Fine-tuning universal machine learning potentials for transition state search in surface catalysis

Raffaele Cheula, Mie Andersen, and John Kitchin

The paper's computational workflow spans two repositories:

| Repository | Purpose | Role in the paper |
|---|---|---|
| ARKIMEDE (arkimede) | <ul><li>Transition state (TS) search workflows</li><li>Active learning loops</li></ul> | <ul><li>Contains BA-Sella TS-search method</li><li>Runs TS-search workflows</li><li>Orchestrates active learning loops</li></ul> |
| MLPs fine-tuning (mlps_finetuning) | <ul><li>Set up of universal machine-learning potentials (MLPs)</li><li>Fine-tuning of MLPs</li></ul> | <ul><li>Contains tools to set up MLP calculators</li><li>Provides functions for the fine-tuning of MLPs from DFT data</li></ul> |

In short, the ARKIMEDE repository manages the execution of structure optimizations, TS-searches, and active learning workflows, while the MLPs fine-tuning repository provides utilities to set up MLP (and DFT) ASE calculators (via the `get_calculator()` function) and to perform MLP fine-tuning (with functions accessible through the `get_finetune_function()` function). These functions can be directly integrated into the ARKIMEDE workflows, enabling standardized execution of the TS-search and active learning calculations without additional setup.

## Repository 1: ARKIMEDE

**Installation**

```
git clone https://github.com/raffaelecheula/arkimede.git
cd arkimede
pip install -e .
```

**Dependencies**

- Core: Numpy, SciPy, Matplotlib, ASE
- TS-searches: Sella
- SLURM workflow manager: Shephex
- Other: PyYAML, tqdm

**Key modules relevant to the paper**

- `arkimede/workflow/calculations.py`
  Contains functions to launch single-point, relaxation, NEB, dimer, ARPESS, Sella, BA-Sella, IRC, vibrations calculations, and manage those calculations through observers.

- `arkimede/workflow/sella.py`
  Contains functions to estimate the Hessian matrix (from finite differences, from TS mode, and from TS bonds), and the observer that defines the BA-Sella method: `modify_hessian_obs()`. This observer modifies the Hessian during the TS-search with Sella by injecting curvature in the direction of the TS bonds when the TS mode is not aligned with that direction (their dot product is below a threshold value).

- `arkimede/workflow/recipes.py`
  Contains the function that defines the workflow for TS-search from initial and final states structures of a reaction step: `search_TS_from_atoms_IS_and_FS()`. First, the initial and final state structures are relaxed, then the reaction step is approximated with the IDPP method; a following NEB calculation can be run to get an improved TS guess structure; then, the TS-search is run. After the TS-search, the TS structure is relaxed following the two directions of the TS mode, and the connectivity of the obtained structures is checked against the original IS and FS structures (to verify if the TS structure corresponds to the starting reaction step). The stochastic-restart procedure is managed by the `rattle_atoms_and_parameters()` function.

- `arkimede/workflow/active_learning.py`
  Contains the implementation of the active learning workflows (sequential and batch) used to iteratively fine-tune the MLPs during the TS searches. The module manages the interaction between TS-search calculations, ASE database construction, and MLP fine-tuning functions. Sequential active learning loops can be run with the `run_active_learning()` function, while batch active learning workflows can be run with the `run_batch_active_learning()` function.

## Repository 2: MLPs fine-tuning

### Installation

```
git clone https://github.com/raffaelecheula/mlps_finetuning.git
cd mlps_finetuning
pip install -e .
```

### Dependencies

- Core: Numpy, SciPy, Matplotlib, PyTorch, ASE
- MLPs libraries: OCP (FAIRChem version 1), FAIRChem, CHGNet, MACE
- DFT codes: Quantum Espresso, VASP
- Other: PyYAML, tqdm

### Key modules relevant to the paper

- `mlps_finetuning/calculators.py`
  Provides a unified interface to instantiate MLP and DFT calculators. It supports CHGNet, MACE, OCP, and FAIRChem models, as well as Quantum Espresso and VASP backends. The module standardizes calculator setup through dedicated wrapper functions and a general `get_calculator()` function that parses the requested calculator/model name. The module also exposes a `get_finetune_function()` dispatcher that links each supported MLP family to its corresponding fine-tuning routine.

## Examples

The Python scripts to reproduce the results reported in the paper can be found in the `examples` folder of the ARKIMEDE repository.

**1) One TS-search**

Run one TS-search calculation with a universal MLP (eSEN-OAM), at given initial and final state structures of a reaction step, using the `search_TS_from_atoms_IS_and_FS()` function. The calculator and the starting structures can be substituted with user-defined ones, allowing the user to run the workflow with their systems.

**2) TS searches with MLPs**

Run TS-search calculations with a universal MLP (eSEN-OAM) on 250 reaction steps from the input database describing $CO_2$ hydrogenation reaction steps on metal and single-atom alloy surfaces. All the calculations are run using the `search_TS_from_atoms_IS_and_FS()` function. The MLP calculator, the TS-search method, and their parameters can be changed to reproduce the results reported in the paper.

### 3) Optimization with DFT

Run single-structure TS searches (BA-Sella) with DFT (Quantum Espresso), starting from TS guess structures obtained with the IDPP interpolation method.

### 4) Optimization with DFT after MLP

Run single-structure TS searches (BA-Sella) with DFT (Quantum Espresso), starting from TS guess structures obtained with TS searches using a MLP (eSEN-OAM).

### 5) Sequential active learning

Run active learning TS searches, in which the MLP (eSEN-OAM) is iteratively fine-tuned using forces evaluated with DFT (Quantum Espresso) single-point calculations. A different MLP is fine-tuned for each TS structure.

### 6) Batch active learning

Run active learning TS searches, in which the MLP (eSEN-OAM) is iteratively fine-tuned using forces evaluated with DFT (Quantum Espresso) single-point calculations. The same MLP is fine-tuned using data from all TS structures.

## Calculation settings

The following parameters are used in the calculations described in the paper:

| Calculation | Parameter | Value |
|---|---|---|
| Relaxation of IS and FS structures<br>File: `arkimede/workflow/calculations.py`<br>Function: `run_relax_calculation()` | `fmax` | `0.05` |
| | `max_steps` | `500` |
| | `optimizer` | `ase.optimize.BFGS` |
| IDPP interpolation<br>File: `arkimede/workflow/transition_states.py`<br>Function: `get_idpp_interpolated_images()` | `n_images` | `10` |
| | `f_max` | `0.01` |
| | `max_steps` | `1000` |
| | `optimizer` | `ase.optimize.LBFGS` |
| NEB calculation<br>File: `arkimede/workflow/calculations.py`<br>Function: `run_neb_calculation()` | `f_max` | `0.05` |
| | `method` | `"ase-neb"` |
| | `optimizer` | `ase.optimize.BFGS` |
| | `k_neb` | `1.0` |
| Dimer TS-search calculation<br>File: `arkimede/workflow/calculations.py`<br>Function: `run_dimer_calculation()` | `f_max` | `0.05` |
| | `trial_trans_step` | `0.06` |
| | `dimer_separation` | `0.02` |
| ARPESS (ClimbFixInternals) TS-search calculations<br>File: `arkimede/workflow/calculations.py`<br>Function: `run_climbfixint_calculation()` | `f_max` | `0.05` |
| | `maxstep` | `0.05` |
| | `optB_kwargs` | `{"maxstep": 0.05}` |
| Sella TS-search calculation<br>File: `arkimede/workflow/calculations.py`<br>Function: `run_sella_calculation()` | `f_max` | `0.05` |
| | `eta` | `0.01` |
| | `gamma` | `0.05` |
| | `delta0` | `0.05` |
| BA-Sella TS-search calculation<br>File: `arkimede/workflow/calculations.py`<br>Function: `run_sella_calculation()` | `f_max` | `0.05` |
| | `eta` | `0.01` |
| | `gamma` | `0.05` |
| | `delta0` | `0.05` |
| | `dot_prod_thr` | `0.50` |
| Finite-differences Hessian calculation<br>File: `arkimede/workflow/calculations.py`<br>Function: `run_vibrations_calculation()` | `delta` | `0.01` |
| | `nfree` | `2` |
| | `indices_vib` | `"not-fixed"` |

The following parameters are used in the stochastic-restart strategy described in the paper:

| Component | Parameter | Value |
|---|---|---|
| Random translation and rattle of atoms File: `arkimede/workflow/recipes.py` Function: `rattle_atoms_and_parameters()` | `indices_rattle` | `"adsorbate"` |
| | `scale_transl` | 0.25 |
| | `scale_rattle` | 0.15 |
| Parameters ranges for BA-Sella calculation File: `arkimede/workflow/recipes.py` Function: `rattle_parameters()` | `eta` | (0.001, 0.010) |
| | `gamma` | (0.01, 0.10) |
| | `delta0` | (0.01, 0.10) |
| | `dot_prod_thr` | (0.00, 0.80) |